\documentclass[submission, Proceedings]{SciPost}

\hypersetup{
    colorlinks,
    linkcolor={red!50!black},
    citecolor={blue!50!black},
    urlcolor={blue!80!black}
}

\usepackage[bitstream-charter]{mathdesign}
\DeclareSymbolFont{usualmathcal}{OMS}{cmsy}{m}{n}
\DeclareSymbolFontAlphabet{\mathcal}{usualmathcal}

\usepackage{amsmath}
\usepackage{braket}

\begin{document}

\newcommand{\pip}{\pi^+}
\newcommand{\pim}{\pi^-}
\newcommand{\alu}{A_{LU}}
\newcommand{\gperp}{G_1^{\perp}}
\newcommand{\hperp}{H_1^{\perp}}
\newcommand{\hangle}{H_1^{\sphericalangle}}
\newcommand{\gpw}{G_1^{\perp\ket{\ell,m}}}
\newcommand{\hpw}{H_1^{\ket{\ell,m}}}
\newcommand{\phih}{\phi_h}
\newcommand{\phir}{\phi_R}

\begin{center}{\Large \textbf{
Multidimensional partial wave analysis of SIDIS dihadron beam spin asymmetries at CLAS12
}}\end{center}

\begin{center}
C. Dilks\textsuperscript{1*} for the CLAS collaboration
\end{center}

\begin{center}
{\bf 1} Duke University, Durham, NC 27710, USA
\\
* christopher.dilks@duke.edu
\end{center}

\begin{center}
\today
\end{center}


\definecolor{palegray}{gray}{0.95}
\begin{center}
\colorbox{palegray}{
  \begin{tabular}{rr}
  \begin{minipage}{0.1\textwidth}
    \includegraphics[width=22mm]{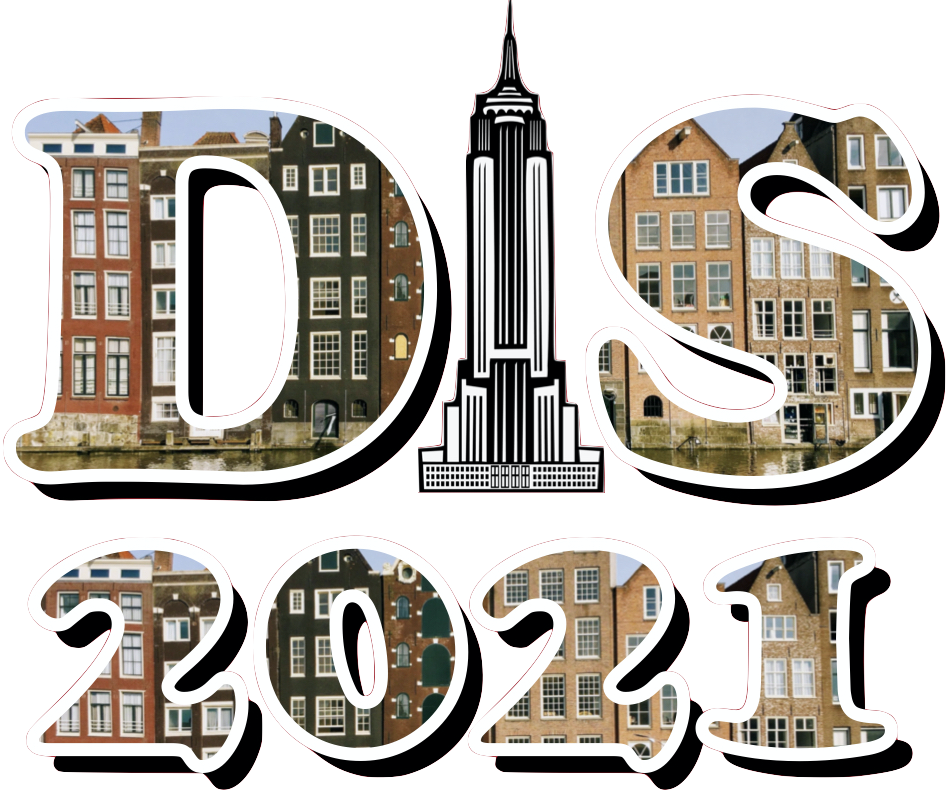}
  \end{minipage}
  &
  \begin{minipage}{0.75\textwidth}
    \begin{center}
    {\it Proceedings for the XXVIII International Workshop\\ on Deep-Inelastic Scattering and
Related Subjects,}\\
    {\it Stony Brook University, New York, USA, 12-16 April 2021} \\
    \doi{10.21468/SciPostPhysProc.?}\\
    \end{center}
  \end{minipage}
\end{tabular}
}
\end{center}

\section*{Abstract}
{\bf
Dihadron beam spin asymmetries provide a wide range of insights into nucleon structure and hadronization. Recent measurements at CLAS12 provide the first empirical evidence of nonzero $G_1^\perp$, the parton helicity-dependent dihadron fragmentation function (DiFF) encoding spin-momentum correlations in hadronization. These measurements also allow for a point-by-point extraction of the subleading-twist PDF $e(x)$ in a collinear framework. We observe different behavior of the asymmetries in different invariant mass regions, motivating a fully multidimensional study. The DiFFs also expand in terms of partial waves, each corresponding to the interference of dihadrons of particular polarizations. Altogether a fully multidimensional partial wave analysis is needed, and this presentation will summarize the efforts and results obtained thus far.
}

\section{Introduction}
\label{sec:intro}
Semi-Inclusive Deep Inelastic Scattering (SIDIS) has proven to be a versatile technique to study the nature of the nucleon. Numerous spin-momentum and spin-spin correlations in hadronization are accessible in dihadron production from polarized SIDIS, which involves the scattering of an electron off a nucleon and the detection of the scattered electron along with a hadron pair, a dihadron~\cite{Bacchetta:2002ux,Bacchetta:2003vn,Gliske:2014wba}. This study focuses on semi-inclusive charged pion pairs from the $e^-p \to e^- \pip\pim X$ process using the upgraded CEBAF Large Acceptance Spectrometer (CLAS12), extending the previous measurement~\cite{Hayward:2021psm} with a partial wave analysis, which explores correlations with dihadron angular momentum.

The SIDIS dihadron-production cross section factorizes into a Parton Distribution Function (PDF), the hard scattering of the electron and quark, and a Dihadron Fragmentation Function (DiFF), which models the formation of two hadrons from the scattered quark. Since the hard scattering cross section is calculable via perturbative QCD and QED, measurements of the SIDIS dihadron cross section constrain the involved PDFs and DiFFs \cite{Aidala:2012mv,Metz:2016swz,Anselmino:2020vlp, Avakian:2019drf}. The presented measurement is of the beam spin asymmetry
\begin{equation}
\alu=\frac{1}{P_B}\frac{N^+-N^-}{N^++N^-},
\end{equation}
where $N^\pm$ is the dihadron yield from scattering a $\pm$ helicity electron, with helicity defined in the proton rest frame, and $P_B$ is the electron beam polarization. Since the spin of the electron is correlated with the spin of the quark, $\alu$ is directly sensitive to spin effects in hadronization, modelled by spin-dependent DiFFs.

There are three types of DiFFs at leading twist, analogous to the three collinear PDFs: the unpolarized $D_1$, the quark longitudinal spin-dependent $\gperp$, and the transverse spin-dependent $\hperp$ and $\hangle$. $\alu$ contains several modulations of the dihadron angles $\phih$, $\phir$, and $\theta$, and the corresponding amplitudes are sensitive to different combinations of PDFs and DiFFs~\cite{Bacchetta:2002ux,Bacchetta:2003vn}. Although transverse momentum-dependent factorization of the cross section at subleading twist is not yet proven~\cite{Bacchetta:2019qkv}, $Q^2$ at CLAS12 reaches down to 1~GeV$^2$, where twist-3 effects are non-negligible. Collinear twist-3 PDFs are significant~\cite{Courtoy:2014ixa}, however subleading-twist DiFFs are likely subdominant~\cite{Yang:2019aan,Pereira:2014hfa,Sirtl:2017rhi}.

Dependence on the angular momentum of the dihadron can be included in DiFFs via a partial wave expansion. Each term describes the interference of dihadrons of particular polarizations, for example an $sp$-wave involves the interference of an unpolarized $s$-state dihadron with a longitudinally or transversely polarized $p$-state dihadron. The partial waves are denoted by angular momentum eigenstates $\ket{\ell,m}$, where $\ell=0,1,2$ respectively corresponds to the $ss,sp,pp$ partial waves, and $m$ further enumerates the polarization possibilities \cite{Gliske:2014wba}. Higher values of $\ell$ are likely suppressed for the relevant mass range. Each modulation amplitude of $\alu$ corresponds to a particular partial wave term of the DiFFs, denoted by $\gpw$ and $\hpw$. The presented $\alu$ measurement is focused on providing access to these partial waves, which allow insight into correlations in hadronization between the fragmenting quark spin and the dihadron angular momentum. 

At twist-2, $\alu$ is sensitive to the well-constrained unpolarized PDF $f_1$ convolved with $\gpw$~\cite{Matevosyan:2017liq}. With $\gpw$ chiral-even, $G_1^{\perp\ket{\ell,-m}}=G_1^{\perp\ket{\ell,m}}$ and $G_1^{\perp\ket{\ell,0}}=0$, therefore only the three amplitudes $\ket{1,1}$, $\ket{2,1}$, and $\ket{2,2}$ are studied. The twist-3 terms of $\alu$ involve the collinear twist-3 PDF $e(x)$ coupled with $\hpw$~\cite{Bacchetta:2003vn}. The recent $\alu$ measurement~\cite{Hayward:2021psm} allows for a point-by-point extraction of $e(x)$ \cite{Courtoy:2014ixa}, given constraints on $\hangle$~\cite{Vossen:2011fk,Courtoy:2012ry}. With $e(x)$ in hand, the presented measurement can constrain the partial wave terms of $\hpw$. Since $\hpw$ is chiral-odd, all nine amplitudes up to $\ell=2$ are measured. 

\section{Experiment and Analysis}
\label{sec:analysis}
The measurement was performed at CLAS12, the upgrade of CLAS, at Jefferson Lab~\cite{Burkert:2020akg}. A longitudinally polarized electron beam was scattered off protons within a liquid hydrogen target. Data from a 10.6 GeV electron beam from Fall 2018, which were used in \cite{Hayward:2021psm} with an approximate integrated charge of 35.7 mC, were combined with data from a 10.2 GeV beam from Spring 2019 (46.8 mC), with the toroid polarity such that electrons bend inward.

The event selection criteria are the same as the previous measurement \cite{Hayward:2021psm}. Only events with $Q^2>1~\text{GeV}^2$, $W>2$~GeV, and $y<0.8$ were analyzed. The pions were required to have positive Feynman-$x$ ($x_F>0$) and a minimum momentum of 1.25~GeV, and the dihadron a maximum $z$ of 0.95. Exclusive reactions were suppressed with a missing mass cut of $M_X>1.5$~GeV. The electrons and dihadrons were detected in the forward detector region with polar scattering angle between $5^\circ$ and $35^\circ$ along with additional fiducial volume cuts, vertex cuts, and particle identification refinements.

Acceptance limitations cause the amplitudes to be linearly dependent, therefore all twelve amplitudes must be fitted for simultaneously. The twist-3 amplitudes with $m=0$ have large fit uncertainty, since $\ket{0,0}$ and $\ket{2,0}$ are highly correlated and $\ket{1,0}$ is suppressed by the phase space; although they are excluded from the figures below, they are still included in the fit. The simultaneous fit is performed with an unbinned maximum likelihood method. The large number of parameters motivates Markov Chain Monte Carlo (MCMC) optimization for the fit, with the likelihood sequentially sampled by the Metropolis-Hastings algorithm~\cite{van2018simple,Bonamente2017,metropolis,hastings} with a Gaussian proposal function.

\section{Results}
\label{sec:results}
\begin{figure}[t]
\centering
\includegraphics[width=0.5\textwidth]{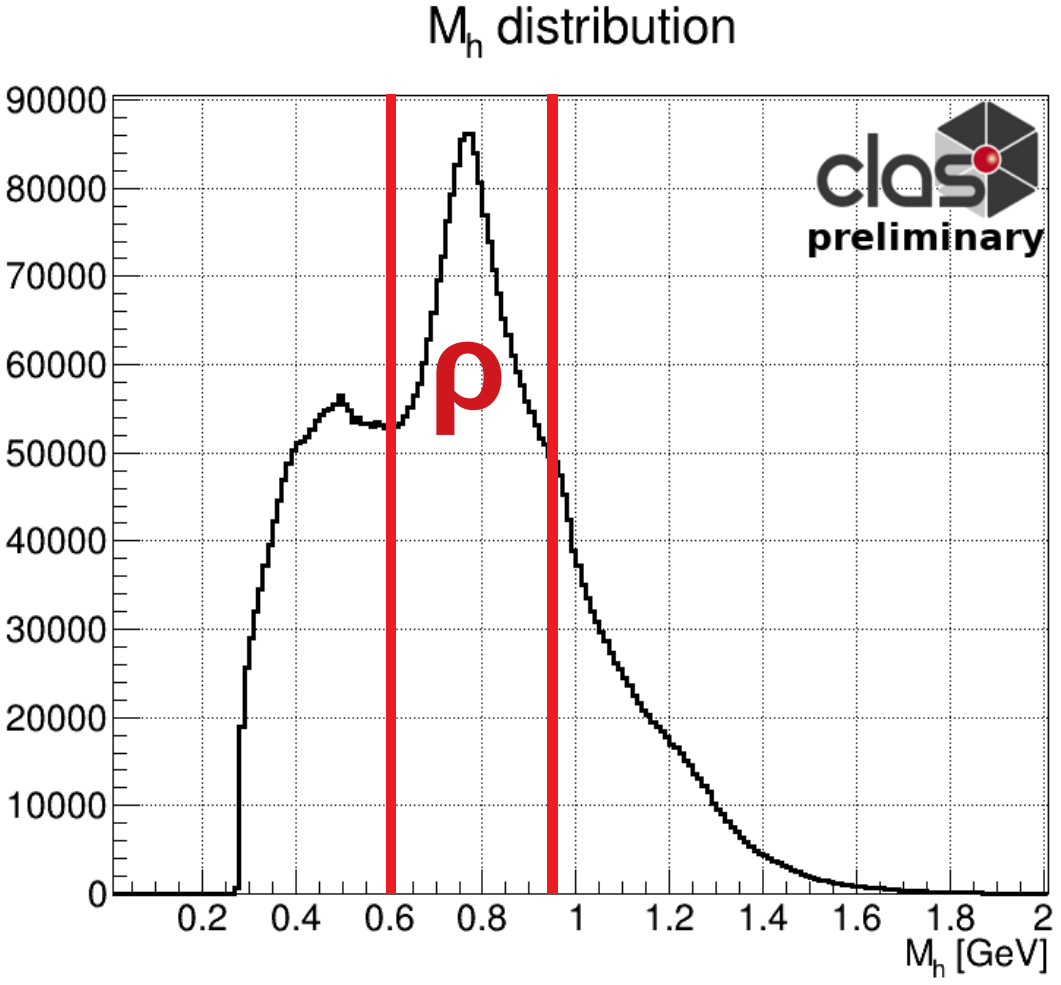}
\caption{Invariant mass $M_h$ distribution of $\pip\pim$ dihadrons, with the $\rho^0$ peak labelled. Vertical lines denote bin boundaries at $0.6$~GeV and $0.95$~GeV.}
\label{fig:mhdist}
\end{figure}

Figure~\ref{fig:mhdist} shows a distribution of the invariant mass $M_h$ of the dihadrons. A prominent peak at 0.77 GeV from $\rho^0\to\pi^+\pi^-$ decays is visible, the presence of which motivates separation of the $\alu$ measurement into three $M_h$ bins, indicated by the vertical lines: below, around, and above the $\rho^0$ mass.

\begin{figure}[p]
\centering
\includegraphics[width=0.6\textwidth]{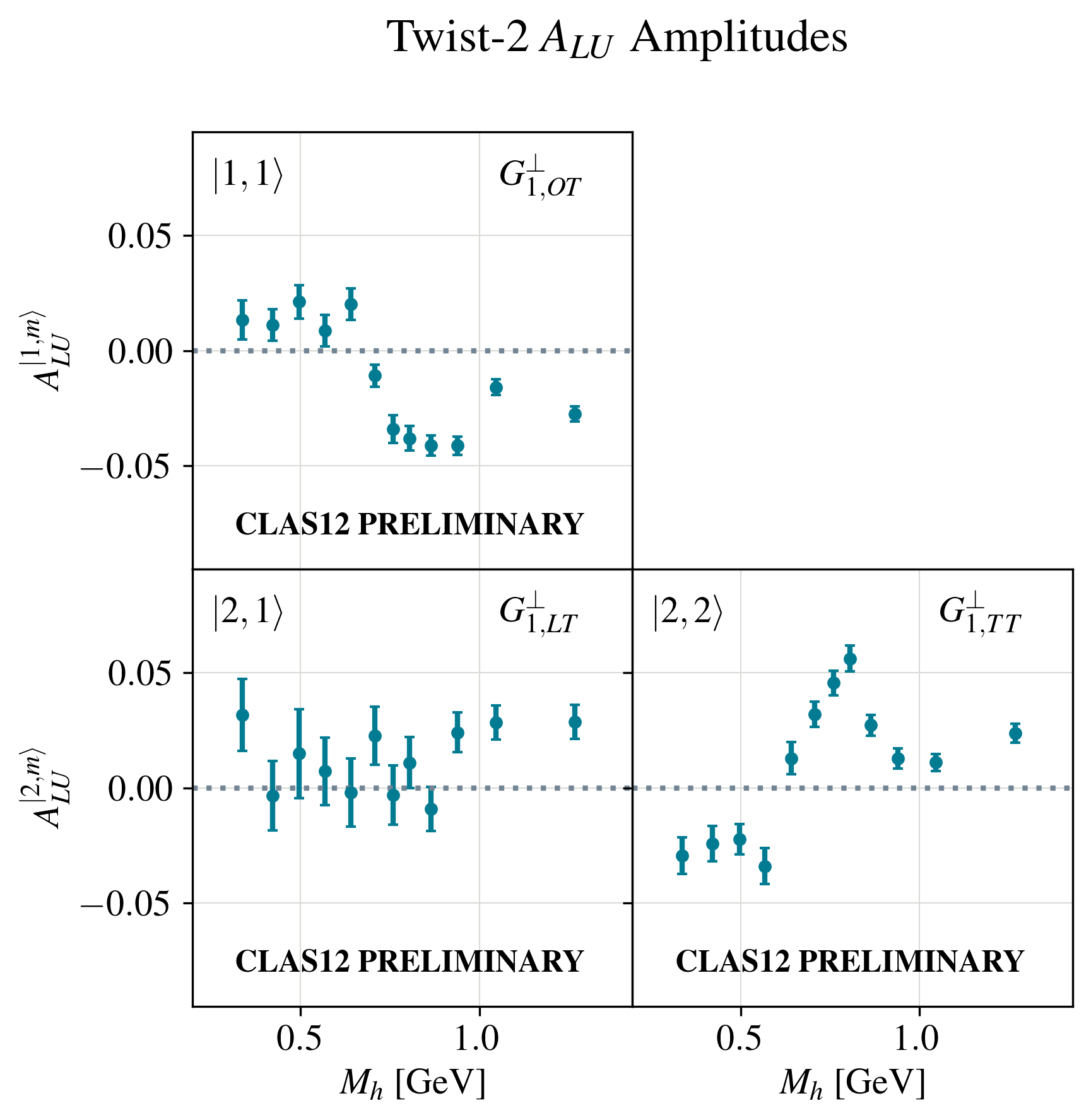}
\caption{Leading-twist $\alu$ amplitudes, in bins of $M_h$, sensitive to $\gpw$.}
\label{fig:tw2mh}
~\\
\includegraphics[width=0.6\textwidth]{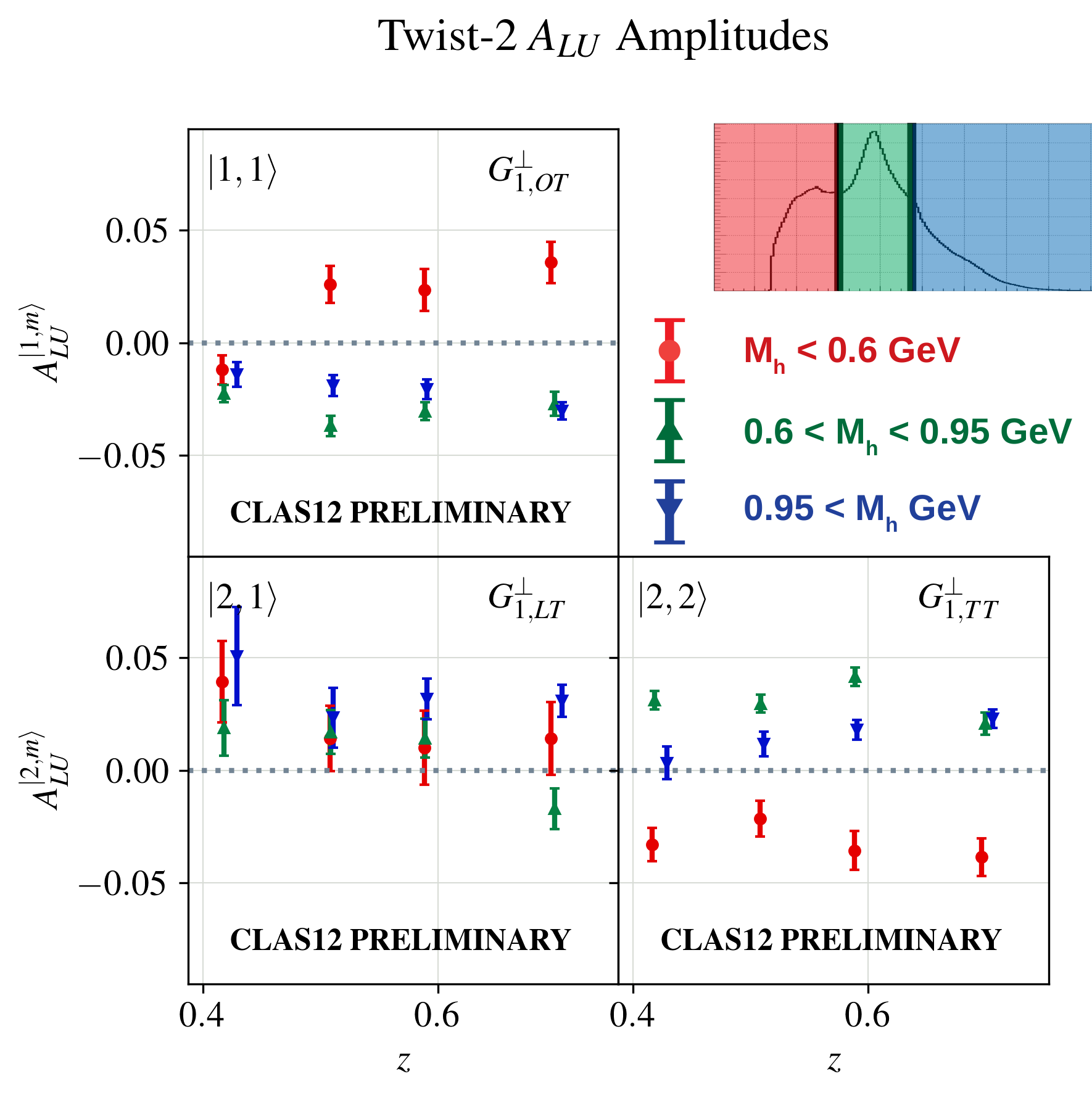}
\caption{Leading-twist $\alu$ amplitudes, in bins of $z$, for three regions of $M_h$: red circles for $M_h<0.6$ GeV, green upward triangles for $0.6<M_h<0.95$ GeV, and blue downward triangles for $M_h>0.95$ GeV. These are sensitive to $\gpw$.}
\label{fig:tw2z}
\end{figure}

\begin{figure}[p]
\centering
\includegraphics[width=\textwidth]{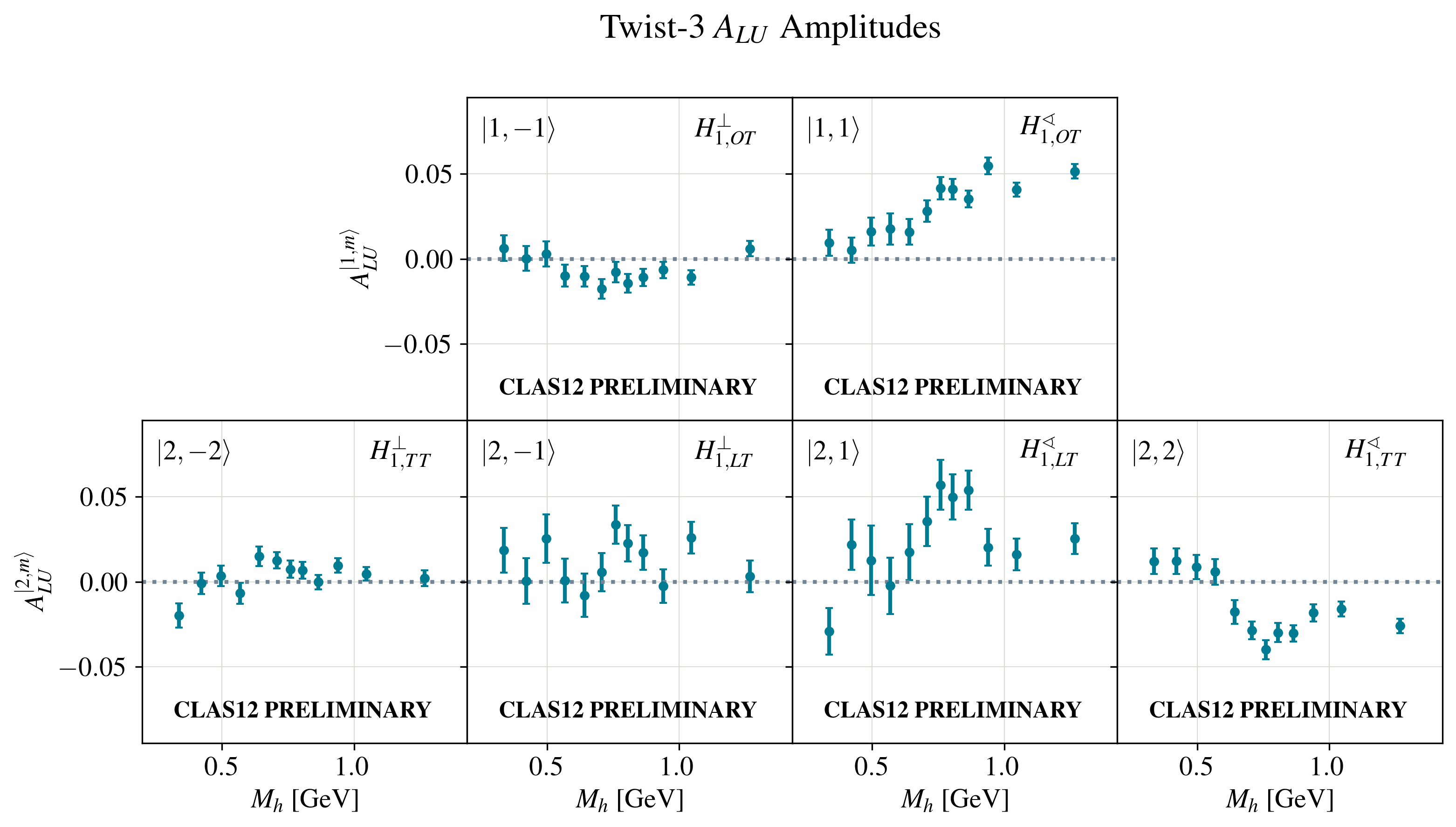}
\caption{Twist-3 $\alu$ amplitudes, in bins of $M_h$, sensitive to $\hpw$.}
\label{fig:tw3mh}
~\\
\includegraphics[width=\textwidth]{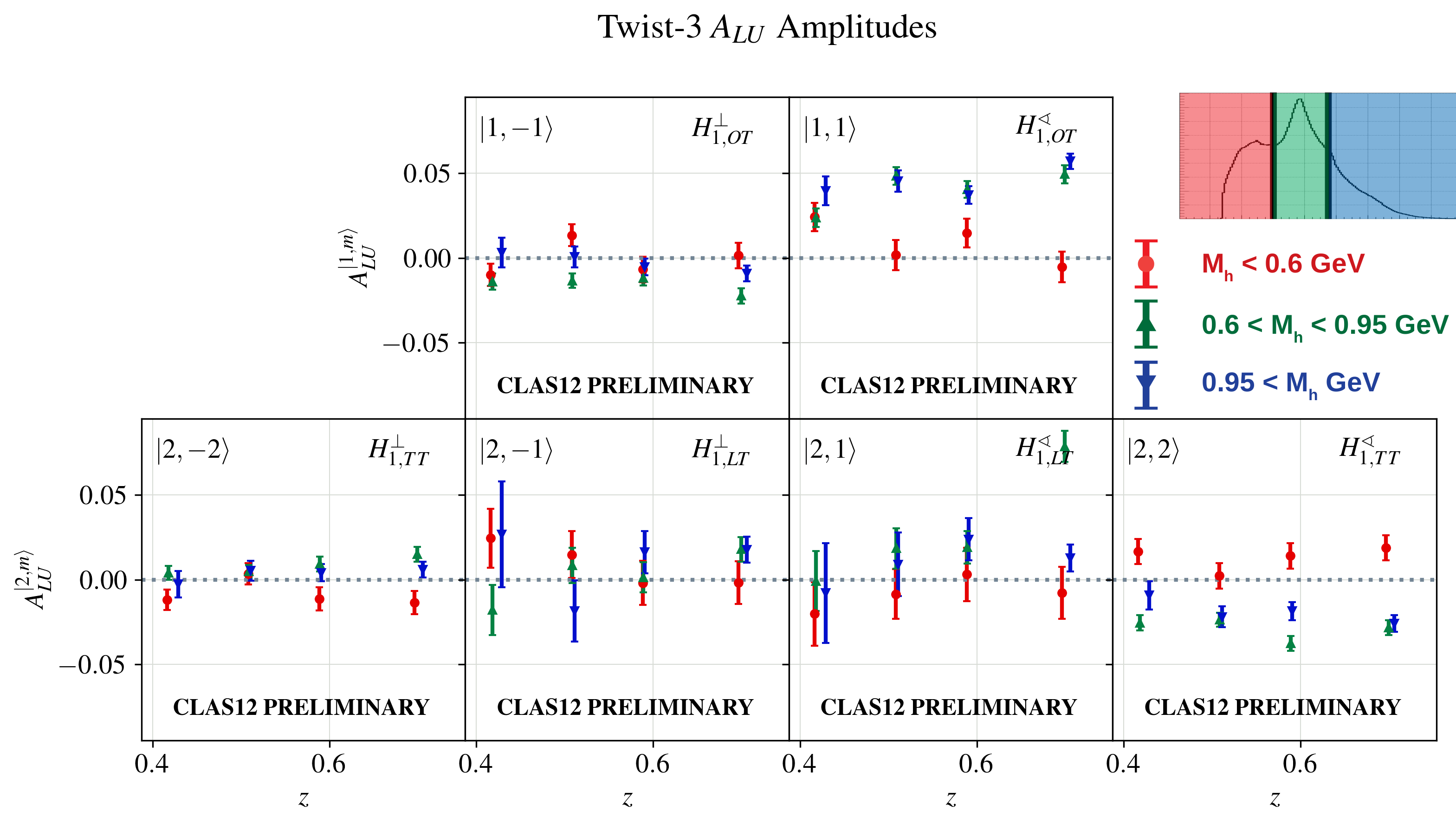}
\caption{Twist-3 $\alu$ amplitudes, in bins of $z$, for three regions of $M_h$: red circles for $M_h<0.6$ GeV, green upward triangles for $0.6<M_h<0.95$ GeV, and blue downward triangles for $M_h>0.95$ GeV. These are sensitive to $\hpw$.}
\label{fig:tw3z}
\end{figure}

The $\alu$ measurements in bins of dihadron invariant mass $M_h$ and $z$ are shown in figures~\ref{fig:tw2mh}-\ref{fig:tw3z}; for the full set of results including other binning schemes, refer to the presentation slides. The rows of subplots of each figure are for values of $\ell$, and the columns for values of $m$. The label in the top-left corner of each subplot gives the partial wave eigenvalues, and the label in the top-right gives the DiFF to which the amplitude is sensitive, including subscripts for the dihadron polarizations~\cite{Gliske:2014wba}: $O$ is for unpolarized, and $L$ and $T$ are for longitudinal and transverse polarization.

The leading-twist amplitudes in 12 bins of $M_h$ are shown in figure~\ref{fig:tw2mh}. Similar to~\cite{Hayward:2021psm}, a sign change is observed near the $\rho^0$ mass. An enhancement at the $\rho^0$ mass is also seen in the $\ket{2,2}$ amplitude, which corresponds to the interference of transversely polarized dihadrons. This enhancement is expected and typical of some $\ell=2$ amplitudes. Since DiFFs are dependent on $z$ in addition to $M_h$, figure~\ref{fig:tw2z} provides the $z$ dependence, in the three regions of $M_h$ described in figure~\ref{fig:mhdist}. Different behavior is clear for the regions $M_h<0.6$ GeV and $M_h>0.6$ GeV, and the amplitude magnitude tends to rise slightly as a function of $z$ and $p_T$.

The twist-3 amplitudes are shown in bins of $M_h$ in figure~\ref{fig:tw3mh}. The $\ket{1,1}$ amplitude, corresponding to $sp$ interference, rises as a function of $M_h$, while the $\ket{2,1}$ and $\ket{2,2}$ amplitudes, corresponding to $pp$ interference, show $\rho^0$ enhancements. Figure~\ref{fig:tw3z} shows the $z$ dependence, for the three $M_h$ regions. Again, the general trend is a slight rise in magnitude with respect to $z$. Interestingly, the $0.6<M_h<0.95$ GeV bin has significant, nonzero $\ket{1,-1}$ and $\ket{2,-2}$ amplitudes, especially at high $z$; the corresponding DiFFs to which these are sensitive involve correlations with the fragmenting quark transverse momentum, and so far have not yet been constrained.

\section{Conclusion}
Measurements of the beam spin asymmetry $\alu$ in SIDIS dihadron production are sensitive to dihadron fragmentation functions $\gperp$ and $H_1$, dependent respectively on the longitudinal and transverse spin of the fragmenting quark. These functions expand in partial waves, each sensitive to the interference of dihadrons with particular polarizations. The presented measurement of partial wave amplitudes of $\alu$ refine access to $\gperp$ extending the prior measurement \cite{Hayward:2021psm} and allow for a better understanding of $\hangle$ and $\hperp$. Future measurements can include different dihadron channels, such as $K\pi$ and $\pi^0\pi^\pm$, which will probe different sets of fragmentation functions, as well as SIDIS from a deuteron target, which will help disentangle flavor dependence.

\section*{Acknowledgements}
We acknowledge the outstanding efforts of the staff of the Accelerator and the Physics Divisions at Jefferson Lab in making this experiment possible.

\paragraph{Funding information}
This work was supported in part by the U.S. Department of Energy, the National Science Foundation (NSF), the Italian Istituto Nazionale di Fisica Nucleare (INFN), the French Centre National de la Recherche Scientifique (CNRS), the French Commissariat pour l$^{\prime}$Energie Atomique, the UK Science and Technology Facilities Council, the National Research Foundation (NRF) of Korea, the Helmholtz-Forschungsakademie Hessen für FAIR (HFHF) and the Ministry of Science and Higher Education of the Russian Federation. The Southeastern Universities Research Association (SURA) operates the Thomas Jefferson National Accelerator Facility for the U.S. Department of Energy under Contract No. DE-AC05-06OR23177.
The work of CD is supported by the U.S. Department of Energy, Office of Science, Office of Nuclear Physics under Award Number DE-SC0019230. 

\bibliography{sources.bib}

\nolinenumbers

\end{document}